\documentclass[twocolumn,showpacs,preprintnumbers,amsmath,amssymb]{revtex4}

\usepackage{epsfig}
\usepackage{graphics}
\usepackage{graphicx}
\usepackage{dcolumn}
\usepackage{bm}

\begin{document}

\preprint{APS/123-QED}

\title{Stability of synchronization in coupled time-delay systems using
Krasovskii-Lyapunov theory}

\author{D.~V.~Senthilkumar$^{1,2}$}
\email{skumar@cnld.bdu.ac.in}
\author{J. Kurths$^{2,3}$}%
 \email{Juergen.Kurths@pik-potsdam.de}
\author{M.~Lakshmanan$^4$}%
 \email{lakshman@cnld.bdu.ac.in}
\affiliation{%
$^1$Centre for Dynamics of Complex Systems, 14469 Potsdam Germany\\
}%
\affiliation{%
$^2$Potsdam  Institute for Climate Impact Research, 14473 Potsdam Germany\\
}%
\affiliation{%
$^3$Institute for Physics, Humboldt University, 12489 Berlin, Germany\\
}%
\affiliation{%
$^4$Centre for Nonlinear Dynamics,Department of Physics,
Bharathidasan University, Tiruchirapalli - 620 024, India\\
}%

\date{\today}

\begin{abstract}

Stability of synchronization in unidirectionally coupled time-delay systems is studied
using the Krasovskii-Lyapunov theory. We have shown that the same general stability condition
is valid for different cases, even for the general situation (but with a constraint) where all the coefficients of the error equation 
corresponding to the synchronization manifold are time-dependent. These analytical
results are also confirmed by numerical simulation of paradigmatic examples.

\end{abstract}

\maketitle


Synchronization of coupled chaotic dynamical systems is an active area of research 
in different branches of science and technology~\cite{asp2001,sbjk2002}. Different 
kinds of chaos synchronization have been reported both theoretically and experimentally 
since its discovery in coupled chaotic oscillators~\cite{hfty1983}. 
Recent studies on
synchronization have been focused on  coupled time-delay systems with or without time-delay 
coupling  because of its intrinsic nature of generating high dimensional chaotic signals~\cite{jdf1982} and
the ease of experimental realization of these systems~\cite{ankp1995}. These systems have potential
applications in secure communication, cryptography, controlling, long term prediction,
optimization of nonlinear system performance, modelling brain activity, 
pattern recognition,  etc.~(cf.~\cite{jdf1982,ankp1995,nnk1963,kihd1980,gdvrr1998,aads2005,huv2002}). 
Specifically,  exploiting synchronization,
communication with chaotic lasers was demonstrated in~\cite{gdvrr1998} and digital information
at gigabyte rates was transferred successfully~\cite{aads2005}  exploiting time-delayed feedback
to generate high-dimensional, high-capacity waveforms at high bandwidths.
With this kind of applications of high-dimensional chaotic signals of time-delay systems,
it becomes extremely important to establish conditions under which synchronized states are
asymptotically stable in coupled  time-delay systems.  For this purpose, the Krasvoskii-Lyapunov
theory has become an extremely useful tool~\cite{nnk1963,huv2002,szhl2007,dvskml2005,dvskjk2009}.

On the other hand, recently, it has been pointed out that the Krasovskii-Lyapunov
theory is not suitable for a more general case where the error equation corresponding to
the synchronization manifold is time-dependent; especially when all its coefficients
are time-dependent~\cite{szhl2007}. In this contribution 
we will show that
the Krasovskii-Lyapunov theory  is not restricted to rather special cases discussed 
in the literature so far, but that it can be exploited as a powerful tool in identifying the
synchronization thresholds and the stability of synchronization in  general coupled time-delay systems.
In particular, we will show that the same
general stability condition resulting from the Krasovskii-Lyapunov 
theory  is indeed valid for  the  general case, 
where the coefficients are time dependent. This general situation is very important in
many applications such as synchronization via dynamical relaying~\cite{ifrv2006}, for instance, synchronization of neural activity in brain with complex functional architecture has been shown to underlie cognitive acts~\cite{erng1999}, 
in dynamically evolving networks~\cite{aaad2008} such as ad-hoc networks~\cite{rh2006}, etc. 
In particular,  we will discuss all the four possible cases that arise
due to the nature of the coefficients in the error equation
corresponding to the synchronization manifold and show that the same stability condition
deduced from the Krasovskii-Lyapunov functional approach is valid for all the cases,
subject to certain conditions.
We will also confirm these analytical results by numerical analysis using paradigmatic
examples. 


Consider the following linearly coupled scalar time-delay system,
\begin{subequations}
\begin{eqnarray}
\dot{x}(t)&=&-ax(t)+bf(x(t-\tau)),  \\
\dot{y}(t)&=&-ay(t)+bf(y(t-\tau))+K(t)(x(t)-y(t)),
\end{eqnarray}
\label{eq.one}
\end{subequations}
where $a$ and $b$ are positive constants, $\tau>0$ is the delay-time, $K(t)$
is the coupling function between the drive and the response systems and $f(x)$ is 
some nonlinear function. Now we can deduce the stability condition for complete synchronization
of the general unidirectionally coupled time-delay systems (\ref{eq.one}). The
time evolution of the difference system with the state variable $\Delta=x(t)-y(t)$
(the error equation corresponding to the complete synchronization manifold 
of the coupled time-delay system (\ref{eq.one})) for small values of $\Delta$ 
can be written as
\begin{align}
\dot{\Delta}=-(a+K(t))\Delta+bf^\prime(y(t-\tau))\Delta_{\tau},\;\;\Delta_\tau=\Delta(t-\tau).
\label{eq.difsys}
\end{align}
It is to be noted that there arises four cases depending on the nature of the
coefficient of the $\Delta$ and $\Delta_\tau$ terms of the above error equation
as follows:
\begin{enumerate}
\item Both coefficients of the $\Delta$ and $\Delta_\tau$ terms are 
time-independent.
\item The coefficient of the $\Delta$ term is time-independent and that of the
$\Delta_\tau$ term is time-dependent.
\item The coefficient of the $\Delta$ term is time-dependent and that of the
$\Delta_\tau$ term is time-independent.
\item Both coefficients of the $\Delta$ and $\Delta_\tau$ terms are 
time-dependent.
\end{enumerate}

The synchronization manifold of the error equation (\ref{eq.difsys}) is locally 
attracting if the origin of this equation is stable. Following the Krasovskii-Lyapunov
theory~\cite{nnk1963}, we define a continuous, positive-definite Lyapunov functional of the form
\begin{align}
V(t)=\frac{1}{2}\Delta^2+\mu\int_{-\tau}^0\Delta^2(t+\theta)d\theta,\qquad V(0)=0
\label{eq.klf}
\end{align}
where $\mu$  is an arbitrary positive parameter, $\mu>0$.  
The derivative of the functional $V(t)$ along the trajectory of the error 
equation (\ref{eq.difsys}),
\begin{align}
\frac{dV}{dt}=-(a+K(t))\Delta^2+bf^{\prime}(y(t-\tau))\Delta
\Delta_{\tau}+\mu\Delta^2-\mu\Delta_{\tau}^2,
\end{align}
has to be negative to ensure the stability of the solution $\Delta=0$. The requirement
that  $\frac{dV}{dt}<0$ for all $\Delta$ and $\Delta_\tau$,
results in the condition for stability as
\begin{align}
a+K(t)>\frac{b^2}{4\mu}f^{\prime}(y(t-\tau))^2+\mu = \Phi(\mu).
\label{eq.ineq}
\end{align}
Now, $\Phi(\mu)$ as a function of $\mu$ for a given $f^{\prime}(x)$ has an
absolute minimum at 
\begin{align}
\mu=(|bf^{\prime}(y(t-\tau))|)/2, 
\label{muone}
\end{align}
with 
$\Phi_{min}=|bf^{\prime}(y(t-\tau))|$.  Since $\Phi\ge\Phi_{min}=
|bf^{\prime}(y(t-\tau))|$, from the inequality (\ref{eq.ineq}), it turns out that
a sufficient condition for asymptotic stability is
\begin{align}
a+K(t)>|bf^{\prime}(y(t-\tau))|.
\label{eq.asystab}  
\end{align}

It is to be noted that since $\mu$ is an arbitrary positive parameter due to the definition of
the positive definite Lyapunov function (\ref{eq.klf}), the above stability condition is valid only when
$\mu=(|bf^{\prime}(y(t-\tau))|)/2$ is a constant, i. e., only when $f^{\prime}(x)$ 
is a constant (in other words when the coefficient of $\Delta_\tau$ term in the
error equation (\ref{eq.difsys}) is time-independent, which corresponds to the cases ($1$) and
($3$) discussed above). On the other hand if $f^{\prime}(x)$ 
is time-dependent, then $\mu$ can
be obtained alternatively by rewriting Eq.~(\ref{eq.ineq}) as
\begin{eqnarray}
b^2f^{\prime}(y(t-\tau))^2&<&4\mu(a+K(t)-\mu),\\ 
                            &=&-4\left[\mu-(a+K(t))/2\right]^2 \nonumber
                            +(a+K(t))^2,\\ \nonumber
                       &\equiv &\Psi(\mu). \nonumber
\label{eqabcd}
\end{eqnarray}
Now, $\Psi(\mu)$ as a function of $\mu$ for a given $f^{\prime}(x)$ has an
absolute maximum at 
\begin{align}
\mu=(a+K(t))/2, 
\label{mutwo}
\end{align}
with 
$\Psi_{max}=(a+K(t))^2$. 
Using this maximum value in the 
right hand side of ($8$), we obtain the same stability condition as that 
of (\ref{eq.asystab}), provided $(a+K(t))/2 > 0 $ since $\mu > 0$. Since $a>0$, 
this implies $K(t) > -a$, that is coupling function $K(t)$ should be either positive
definite or $|K(t)| > a$ 
if it is negative. In particular for the case 2, since the coefficient of the 
$\Delta$ term in the error equation is time independent (which 
corresponds to the cases ($1$) and ($2$) mentioned above), 
$K(t) = k > -a$ for all $t$ ($k:const.$).

However, there
arises an even more general situation where the coefficients of both the $\Delta$
and $\Delta_\tau$ terms are time dependent (case $4$), in which case the arbitrary positive 
parameter $\mu$ in the Lyapunov functional has to be
 chosen as a positive definite function, $\mu=g(t)>0$ for all $t$. In this case, one has to consider the 
derivative of $\mu=g(t)$ also in the derivative of $V(t)$ as follows,
\begin{eqnarray}
\frac{dV}{dt}&=&-(a+K(t))\Delta^2+bf^{\prime}(y(t-\tau))\Delta
\Delta_{\tau}\\ \nonumber
&+&g(t)(\Delta^2-\Delta_{\tau}^2)+\dot{g}(t)\int_{-\tau}^0\Delta^2(t+\theta)d\theta<0.
\end{eqnarray}
It is known from the Lyapunov functional that the term 
$\int_{-\tau}^0\Delta^2(t+\theta)d\theta$ is positive and let us suppose that
$\dot{g}(t)\le 0$ for all $t$, then for $\dot{V}(t)< 0$ a sufficient condition is that
\begin{eqnarray}
-(a+K(t))\Delta^2+bf^{\prime}(y(t-\tau))\Delta
\Delta_{\tau}\\ \nonumber
+g(t)(\Delta^2-\Delta_{\tau}^2)<0,
\end{eqnarray}
\begin{eqnarray}
-\left[(a+K(t))-b^2f^{\prime}(y(t-\tau))^2/4g(t)-g(t)\right]\Delta^2 \\ \nonumber
-g(t)\left[\Delta_{\tau}-bf^{\prime}(y(t-\tau))\Delta/2g(t)\right]^2<0.
\end{eqnarray}
The second term in the above equation is negative by assumption of $g(t)$ and hence
it follows that
\begin{eqnarray}
b^2f^{\prime}(y(t-\tau))^2&<&4g(t)(a+K(t)-g(t)),\\ \nonumber
                            &=&-4\left[g(t)-(a+K(t))/2\right]^2
                            +(a+K(t))^2, \\ \nonumber
                            &\equiv &\Gamma(g(t)).
\label{eq.ineqthree}
\end{eqnarray}
Consequently we obtain the same stability condition as in Eq.~(\ref{eq.asystab})
with the maximum of $\Gamma$,  $\Gamma_{max}=(a+K(t))^2$, occuring at $g(t)=(a+K(t))/2>0$, 
along with the condition $\dot{g}(t)\le 0$, that is $\dot{K}(t)=dK(t)/dt\le 0$, for all $t$. 

Note that our above analysis holds good in case ($4$), only for $\dot{K}(t)\le 0$ and is
not valid for $\dot{K}(t)>0$.  For the latter case,
we are unable to obtain a sufficiency condition yet.  Consequently the cases ($1$)-($3$) 
cannot be treated as special cases of the most general case ($4$) at present.

Thus, we have shown that the same general stability condition, Eq.~(\ref{eq.asystab}), is valid for
all the four cases that arise in the error equation (\ref{eq.difsys}) corresponding 
to the synchronization manifold of the unidirectionally coupled time-delay systems with
a restriction in case ($4$). 

\begin{figure}
\centering
\includegraphics[width=1.0\columnwidth]{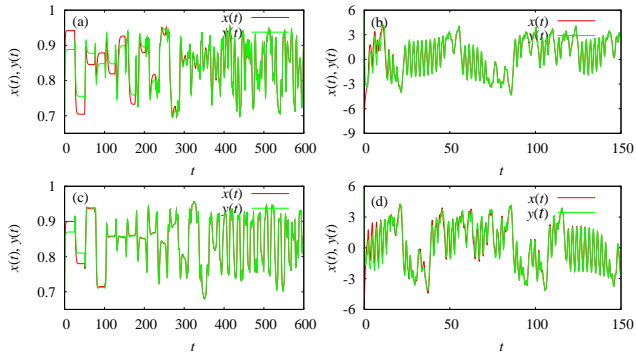}
\caption{\label{fig1} (Color online) The time trajectory plot of the variables
$x(t)$ and $y(t)$ of the coupled time-delay systems (\ref{eq.one}) indicating 
complete synchronization between them. (a) Piecewise linear time-delay system,
(\ref{eq.one}) and (\ref{eqoneb}), for the parameters $a=1.0, b=1.2, \tau=25.0$
and for the constant coupling $k_1=k_2=0.9$. (b) Ikeda time-delay system,
(\ref{eq.one}) and (\ref{eqonec}), for the parameters $a=1.0, b=5, \tau=2.0$ along
with the constant coupling $k_1=k_2=5.0$. (c) Piecewise linear time-delay system for the
same values of the system parameters as in Fig.~\ref{fig1}a with the square wave coupling
rate $k_1=0.9$ and $k_2=1.0$. (d) Ikeda system for the same values of the system
parameters  and with the square wave coupling rate $k_1=5.0$ and $k_2=6.0$.}
\end{figure}

In this section, we will provide numerical confirmation of the above stability
analysis for all the four cases using appropriate nonlinear functional forms 
$f(x)$ and suitable coupling $K(t)$ in the coupled time-delay systems (\ref{eq.one}).  
For this purpose we will consider the nonlinear functions $f(x)$ as the piecewise linear 
function, which has been studied in detail recently~\cite{dvskmlijbc,dvskml2005,dvskml2006},
\begin{eqnarray}
f(x)=
\left\{
\begin{array}{cc}
0,&  x \leq -4/3  \\
            -1.5x-2,&  -4/3 < x \leq -0.8 \\
            x,&    -0.8 < x \leq 0.8 \\              
            -1.5x+2,&   0.8 < x \leq 4/3 \\
            0,&  x > 4/3, \\ 
         \end{array} \right.
\label{eqoneb}
\end{eqnarray}
and 
\begin{equation}
f(x)=\sin(x(t-\tau)),
\label{eqonec}
\end{equation}
which is the paradigmatic Ikeda model~\cite{kihd1980}. We have fixed the
parameters as $a=1.0, b=1.2$ and $\tau=25.0$ for the coupled piecewise linear
time-delay system defined by (\ref{eq.one}) and (\ref{eqoneb}), for which the uncoupled
systems exhibit a hyperchaotic behavior with nine positive Lyapunov exponents
~\cite{dvskmlijbc,dvskml2005,dvskml2006,dvskjk2009}. For the coupled Ikeda systems (\ref{eq.one}) 
and (\ref{eqonec}), the parameters are chosen as
$a=1.0, b=5.0$ and $\tau=2.0$ where the uncoupled individual Ikeda systems exhibit
a hyperchaotic behavior with three positive Lyapunov exponents~\cite{dvskjk2009}.

We choose the coupling function $K(t)$ as a square wave function represented 
as~\cite{mcjk2007}
\begin{equation}
K(t)=\{(t_0,k_1),(t_1,k_2),(t_2,k_1),(t_3,k_2),\ldots\},
\label{eqcoup}
\end{equation}
where $t_j=t_0+(j-1)\tau_s, j\ge 1$ is the switching instant, $k_1>0, k_2>0$ with $k_1\ne k_2$. 
For constant coupling, $K(t)=k_1=k_2$. On the other hand, either $k_1=0$ or
$k_2=0$, then the coupling is called an intermittent coupling/control which is now
being widely studied in the literature~\cite{mz2000}.

First, we use the piecewise linear function  (\ref{eqoneb}),
and the constant coupling $K(t)=k_1=k_2$.  It is clear from the form of the
nonlinear function $f(x)$ and the coupling that both coefficients of the
$\Delta$ and $\Delta_\tau$ terms in the error equation, (\ref{eq.difsys}),
are constant (case $1$) and consequently $\mu$ can either be chosen as $\mu=(|bf^{\prime}(y(t-\tau))|)/2$
or $\mu=(|a+K(t)|)/2$. The time trajectory of the variables $x(t)$ and $y(t)$ of
the coupled piecewise linear time-delay systems, (\ref{eq.one}) and (\ref{eqoneb}), are
shown in Fig.~\ref{fig1}a indicating complete synchronization between them for 
the coupling strength $k=k_1=k_2=0.9$ satisfying the stability condition 
$a+k>bf^{\prime}(y(t-\tau))=1.5b$. Here, the other system parameters are 
fixed as noted above. 

Next, we analyse the function $f(x)=\sin(x(t-\tau))$, given by (\ref{eqonec}), of the 
Ikeda system with constant coupling, which corresponds to the case~$2$ where the
coefficient of the $\Delta_\tau$ term in the error equation is time-dependent, while that of the
$\Delta$ term is still time-independent and hence $\mu$ can take the form $\mu=(a+K(t))/2$
with $K(t)>0$.
The coupling strength is fixed as $k=k_1=k_2=5.0$ such that
the stability condition $a+k>bf^{\prime}(y(t-\tau))=b$ is satisfied. The 
variables $x(t)$ and $y(t)$ of the coupled Ikeda systems, (\ref{eq.one}) and (\ref{eqonec}), 
are plotted as a function of time in  Fig.~\ref{fig1}b demonstrating complete synchronization
between them.

Again, we consider the piecewise linear function  (\ref{eqoneb}), and
the same parameter values as in the case $1$ but with the square wave coupling $K(t)$ chosen
as $k_1=0.9$ and $k_2=1.0$ such that the stability condition (\ref{eq.asystab})
is satisfied for all $t$. The switching instant $\tau_s$  between $k_1$ and $k_2$ for the
square wave coupling rate is fixed as $\tau_s=1.0$. This situation corresponds to the
case~$3$, where the
coefficient of the $\Delta$ term in the error equation is time-dependent, while that 
of the $\Delta_\tau$ term is time-independent and as a result $\mu$ can be 
fixed as $\mu=(|bf^{\prime}(y(t-\tau))|)/2$. The time trajectory of the variables 
$x(t)$ and $y(t)$ are shown in Fig.~\ref{fig1}c indicating complete synchronization.
Note that here $K(t)>0$ and the stability condition (\ref{eq.asystab}) is indeed satisfied.

Finally for the more general case where both coefficients of the $\Delta$ and
$\Delta_\tau$ term of the error equation are time-dependent, $\mu=g(t)$ can be given 
as $g(t)=(a+K(t))/2$ for the chosen form of the square wave coupling $K(t)$ with $K(t)>0$ and
$\dot{K}(t)=0$.
Figure~\ref{fig1}d is plotted for the same values of the system parameters as in Fig.~\ref{fig1}b with
$k_1=5.0, k_2=6.0$ and $\tau_s=1.0$ satisfying the stability condition (\ref{eq.asystab}), indicating
complete synchronization between the variables $x(t)$ and $y(t)$. 

Asymptotic stability of synchronized state in a unidirectionally coupled time-delay
system is studied using the Krasovskii-Lyapunov theory. We have shown that 
the same stability condition is valid for all the four cases that arises due
to the nature of the coefficients of the $\Delta$ and $\Delta_\tau$ terms in
the error equation corresponding to the synchronization manifold. In particular, we
have shown that the same general stability condition is valid even for the 
general case where both coefficients of the $\Delta$ and $\Delta_\tau$ terms
in the error equation are time-dependent,   which is of high
importance for various applications. We have also numerically confirmed these
results using appropriate examples along with suitable coupling configuration.

The work of DVS has been supported by Alexander von Humboldt Foundation.
JK has been supported by his Humboldt-CSIR research award and NoE BIOSIM (EU)
Contract No. LSHB-CT-2004-005137.
ML acknowledges the support from a Department of Science and Technology (DST),
Government of India sponsored IRHPA research project and DST Ramanna Program.


\begin{thebibliography}{42}
\expandafter\ifx\csname natexlab\endcsname\relax\def\natexlab#1{#1}\fi
\expandafter\ifx\csname bibnamefont\endcsname\relax
  \def\bibnamefont#1{#1}\fi
\expandafter\ifx\csname bibfnamefont\endcsname\relax
  \def\bibfnamefont#1{#1}\fi
\expandafter\ifx\csname citenamefont\endcsname\relax
  \def\citenamefont#1{#1}\fi
\expandafter\ifx\csname url\endcsname\relax
  \def\url#1{\texttt{#1}}\fi
\expandafter\ifx\csname urlprefix\endcsname\relax\def\urlprefix{URL }\fi
\providecommand{\bibinfo}[2]{#2}
\providecommand{\eprint}[2][]{\url{#2}}

\bibitem[{\citenamefont{Pikovsky et~al.}(2001)\citenamefont{Pikovsky,
  Rosenbulm, and Kurths}}]{asp2001}
\bibinfo{author}{\bibfnamefont{A.~S.} \bibnamefont{Pikovsky}},
  \bibinfo{author}{\bibfnamefont{M.~G.} \bibnamefont{Rosenbulm}},
  \bibnamefont{and} \bibinfo{author}{\bibfnamefont{J.}~\bibnamefont{Kurths}},
  \emph{\bibinfo{title}{Synchronization - A Unified Approach to Nonlinear
  Science}} (\bibinfo{publisher}{Cambridge University Press},
  \bibinfo{address}{Cambridge}, \bibinfo{year}{2001}).
  
\bibitem[{\citenamefont{Boccaletti et~al.}(2002)\citenamefont{Boccaletti,
  Kurths, Osipov, Valladares, and Zhou}}]{sbjk2002}
\bibinfo{author}{\bibfnamefont{S.}~\bibnamefont{Boccaletti}},
  \bibinfo{author}{\bibfnamefont{J.}~\bibnamefont{Kurths}},
  \bibinfo{author}{\bibfnamefont{G.}~\bibnamefont{Osipov}},
  \bibinfo{author}{\bibfnamefont{D.~L.} \bibnamefont{Valladares}},
  \bibnamefont{and} \bibinfo{author}{\bibfnamefont{C.~S.} \bibnamefont{Zhou}},
  \bibinfo{journal}{Phys. Reports} \textbf{\bibinfo{volume}{366}},
  \bibinfo{pages}{1} (\bibinfo{year}{2002}).

\bibitem[{\citenamefont{Fujisaka and Yamada}(1983)}]{hfty1983}
\bibinfo{author}{\bibfnamefont{H.}~\bibnamefont{Fujisaka}} \bibnamefont{and}
  \bibinfo{author}{\bibfnamefont{T.}~\bibnamefont{Yamada}},
  \bibinfo{journal}{Prog. Theor. Phys.} \textbf{\bibinfo{volume}{69}},
  \bibinfo{pages}{32} (\bibinfo{year}{1983});
\bibinfo{author}{\bibfnamefont{L.~M.} \bibnamefont{Pecora}} \bibnamefont{and}
  \bibinfo{author}{\bibfnamefont{T.~L.} \bibnamefont{Carroll}},
  \bibinfo{journal}{Phys.\ Rev. Lett.} \textbf{\bibinfo{volume}{64}},
  \bibinfo{pages}{821} (\bibinfo{year}{1990});
\bibinfo{author}{\bibfnamefont{J.~F.} \bibnamefont{Heagy}},
  \bibinfo{author}{\bibfnamefont{L.~M.} \bibnamefont{Pecora}},
  \bibnamefont{and} \bibinfo{author}{\bibfnamefont{T.~L.}
  \bibnamefont{Carroll}}, \bibinfo{journal}{Phys. Rev. Lett.}
  \textbf{\bibinfo{volume}{74}}, \bibinfo{pages}{4185} (\bibinfo{year}{1995}).

\bibitem[{\citenamefont{Farmer}(1982)}]{jdf1982}
\bibinfo{author}{\bibfnamefont{J.~D.} \bibnamefont{Farmer}},
  \bibinfo{journal}{Physica D} \textbf{\bibinfo{volume}{4}},
  \bibinfo{pages}{366} (\bibinfo{year}{1982}).

\bibitem[{\citenamefont{Fujisaka and Yamada}(1995)}]{ankp1995}
\bibinfo{author}{\bibfnamefont{A.}~\bibnamefont{Namajunas}}, 
\bibinfo{author}{\bibfnamefont{K.}~\bibnamefont{Pyragas}} \bibnamefont{and}
  \bibinfo{author}{\bibfnamefont{A.}~\bibnamefont{Tamasevicius}},
  \bibinfo{journal}{Phys. Lett. A} \textbf{\bibinfo{volume}{201}},
  \bibinfo{pages}{42} (\bibinfo{year}{1995});
\bibinfo{author}{\bibfnamefont{M. Y.}~\bibnamefont{Kim}}, 
\bibinfo{author}{\bibfnamefont{C.}~\bibnamefont{Sramek}},
\bibinfo{author}{\bibfnamefont{A.}~\bibnamefont{Uchida}} \bibnamefont{and}
  \bibinfo{author}{\bibfnamefont{R.}~\bibnamefont{Roy}},
  \bibinfo{journal}{Phys. Rev. E} \textbf{\bibinfo{volume}{74}},
  \bibinfo{pages}{016211} (\bibinfo{year}{2006});
\bibinfo{author}{\bibfnamefont{S.}~\bibnamefont{Sano}}, 
\bibinfo{author}{\bibfnamefont{A.}~\bibnamefont{Uchida}},
\bibinfo{author}{\bibfnamefont{S.}~\bibnamefont{Yoshimori}} \bibnamefont{and}
  \bibinfo{author}{\bibfnamefont{R.}~\bibnamefont{Roy}},
  \bibinfo{journal}{Phys. Rev. E} \textbf{\bibinfo{volume}{75}},
  \bibinfo{pages}{016207} (\bibinfo{year}{2007}).

\bibitem[{\citenamefont{Ikeda et~al.}(1980)\citenamefont{Ikeda,
  Daido, and Akimoto}}]{kihd1980}
\bibinfo{author}{\bibfnamefont{K.} \bibnamefont{Ikeda}},
  \bibinfo{author}{\bibfnamefont{H.}~\bibnamefont{Daido}},
  \bibnamefont{and} \bibinfo{author}{\bibfnamefont{O.}~\bibnamefont{Akimoto}},
  \bibinfo{journal}{Phys. Rev. Lett.} \textbf{\bibinfo{volume}{45}},
  \bibinfo{pages}{709} (\bibinfo{year}{1980});
\bibinfo{author}{\bibfnamefont{K.} \bibnamefont{Ikeda}}, \bibnamefont{and}
\bibinfo{author}{\bibfnamefont{M.}~\bibnamefont{Matsumoto}},
\bibinfo{journal}{Physica D} \textbf{\bibinfo{volume}{29}},
  \bibinfo{pages}{223} (\bibinfo{year}{1987}).

\bibitem[{\citenamefont{Ikeda et~al.}(1998)\citenamefont{Ikeda,
  Daido, and Akimoto}}]{gdvrr1998}
\bibinfo{author}{\bibfnamefont{G. D.} \bibnamefont{VanWiggeren}},
  \bibnamefont{and} \bibinfo{author}{\bibfnamefont{R.}~\bibnamefont{Roy}},
  \bibinfo{journal}{Science} \textbf{\bibinfo{volume}{279}},
  \bibinfo{pages}{1198} (\bibinfo{year}{1998}).

\bibitem[{\citenamefont{Ikeda et~al.}(2005)\citenamefont{Ikeda,
  Daido, and Akimoto}}]{aads2005}
\bibinfo{author}{\bibfnamefont{A.} \bibnamefont{Argyris}},
\bibinfo{author}{\bibfnamefont{D.} \bibnamefont{Syvridis}},
\bibinfo{author}{\bibfnamefont{L.} \bibnamefont{Larger}},
\bibinfo{author}{\bibfnamefont{V.} \bibnamefont{Annovazzi-Lodi}},
\bibinfo{author}{\bibfnamefont{P.} \bibnamefont{Colet}},
\bibinfo{author}{\bibfnamefont{I.} \bibnamefont{Fischer}},
\bibinfo{author}{\bibfnamefont{J.} \bibnamefont{Garcia-Ojalvo}},
\bibinfo{author}{\bibfnamefont{C. R.} \bibnamefont{Mirasso}},
\bibinfo{author}{\bibfnamefont{L.} \bibnamefont{Pesquera}},
  \bibnamefont{and} \bibinfo{author}{\bibfnamefont{K. A.}~\bibnamefont{Shore}},
  \bibinfo{journal}{Nature (London)} \textbf{\bibinfo{volume}{438}},
  \bibinfo{pages}{343} (\bibinfo{year}{2005}).

\bibitem[{\citenamefont{Krasovskii}(1963)}]{nnk1963}
\bibinfo{author}{\bibfnamefont{N.~N.} \bibnamefont{Krasovskii}},
  \emph{\bibinfo{title}{Stability of Motion}} (\bibinfo{publisher}{Stanford
  University Press}, \bibinfo{address}{Stanford}, \bibinfo{year}{1963});
\bibinfo{author}{\bibfnamefont{Y.} \bibnamefont{Kuang}},
  \emph{\bibinfo{title}{Delay Differential Equations}} (\bibinfo{publisher}{Academic
  Press}, \bibinfo{address}{New York}, \bibinfo{year}{1993});
\bibinfo{author}{\bibfnamefont{J. K.} \bibnamefont{Hale}}\bibnamefont{and}
\bibinfo{author}{\bibfnamefont{S. M. V.} \bibnamefont{Lunel}},
  \emph{\bibinfo{title}{Introduction to Functional Differential Equations}} 
(\bibinfo{publisher}{Springer}, \bibinfo{address}{New York}, \bibinfo{year}{1993});
\bibinfo{author}{\bibfnamefont{K.}~\bibnamefont{Pyragas}},
  \bibinfo{journal}{Phys. Rev. E} \textbf{\bibinfo{volume}{58}},
  \bibinfo{pages}{3067} (\bibinfo{year}{1998}).

\bibitem[{\citenamefont{Voss}(2002)}]{huv2002}
\bibinfo{author}{\bibfnamefont{H.~U.} \bibnamefont{Voss}},
  \bibinfo{journal}{Phys. Rev. E} \textbf{\bibinfo{volume}{61}},
  \bibinfo{pages}{5115} (\bibinfo{year}{2000});
\bibinfo{author}{\bibfnamefont{E.~M.}~\bibnamefont{Shahverdiev}},
  \bibinfo{author}{\bibfnamefont{S.} \bibnamefont{Sivaprakasam}}, \bibnamefont{and}
  \bibinfo{author}{\bibfnamefont{K. A.} \bibnamefont{Shore}},
  \bibinfo{journal}{ibid.} \textbf{\bibinfo{volume}{66}},
  \bibinfo{pages}{017204} (\bibinfo{year}{2002});
\bibinfo{author}{\bibfnamefont{E. M.}~\bibnamefont{Shahverdiev}},
  \bibinfo{journal}{ibid.} \textbf{\bibinfo{volume}{70}},
  \bibinfo{pages}{067202} (\bibinfo{year}{2004}).

\bibitem[{\citenamefont{Boccaletti et~al.}(2001)\citenamefont{Zhou,
  Li, and Wu}}]{szhl2007}
\bibinfo{author}{\bibfnamefont{S.}~\bibnamefont{Zhou}},
  \bibinfo{author}{\bibfnamefont{H.} \bibnamefont{Li}},
  \bibnamefont{and} \bibinfo{author}{\bibfnamefont{Z.}~\bibnamefont{Wu}},
  \bibinfo{journal}{Phys. Rev. E} \textbf{\bibinfo{volume}{75}},
  \bibinfo{pages}{037203} (\bibinfo{year}{2007}).

\bibitem[{\citenamefont{Boccaletti et~al.}(2001)\citenamefont{Zhou,
  Li, and Wu}}]{ifrv2006}
\bibinfo{author}{\bibfnamefont{I.}~\bibnamefont{Fischer}},
  \bibinfo{author}{\bibfnamefont{R.} \bibnamefont{Vicente}},
\bibinfo{author}{\bibfnamefont{J. M.} \bibnamefont{Buldu}},
\bibinfo{author}{\bibfnamefont{M.} \bibnamefont{Peil}},
\bibinfo{author}{\bibfnamefont{C. R.} \bibnamefont{Mirasso}},
\bibinfo{author}{\bibfnamefont{M. C.} \bibnamefont{Torrent}},
  \bibnamefont{and} \bibinfo{author}{\bibfnamefont{J. }~\bibnamefont{Garcia-Ojalvo}},
  \bibinfo{journal}{Phys. Rev. Lett.} \textbf{\bibinfo{volume}{97}},
  \bibinfo{pages}{123902} (\bibinfo{year}{2006}).

\bibitem[{\citenamefont{Boccaletti et~al.}(2001)\citenamefont{Zhou,
  Li, and Wu}}]{erng1999}
\bibinfo{author}{\bibfnamefont{E.}~\bibnamefont{Rodriguez}},
  \bibinfo{author}{\bibfnamefont{N.} \bibnamefont{George}},
\bibinfo{author}{\bibfnamefont{J. -P.} \bibnamefont{Lachaux}},
\bibinfo{author}{\bibfnamefont{J.} \bibnamefont{Martinerie}},
\bibinfo{author}{\bibfnamefont{B.} \bibnamefont{Renault}},
  \bibnamefont{and} \bibinfo{author}{\bibfnamefont{F. J.}~\bibnamefont{Varela}},
  \bibinfo{journal}{Nature (London)} \textbf{\bibinfo{volume}{397}},
  \bibinfo{pages}{430} (\bibinfo{year}{1999}).

\bibitem[{\citenamefont{Boccaletti et~al.}(2001)\citenamefont{Zhou,
  Li, and Wu}}]{aaad2008}
\bibinfo{author}{\bibfnamefont{A.}~\bibnamefont{Arenas}},
  \bibinfo{author}{\bibfnamefont{A.} \bibnamefont{Díaz-Guilera}},
\bibinfo{author}{\bibfnamefont{J.} \bibnamefont{Kurths}},
\bibinfo{author}{\bibfnamefont{Y.} \bibnamefont{Moreno}},
  \bibnamefont{and} \bibinfo{author}{\bibfnamefont{C.}~\bibnamefont{Zhou}},
  \bibinfo{journal}{Physics Reports} \textbf{\bibinfo{volume}{469}},
  \bibinfo{pages}{93} (\bibinfo{year}{2008}).

\bibitem[{\citenamefont{Pikovsky et~al.}(2001)\citenamefont{Pikovsky,
  Rosenbulm, and Kurths}}]{rh2006}
\bibinfo{author}{\bibfnamefont{R.} \bibnamefont{Hekmat}},
  \emph{\bibinfo{title}{Ad-hoc Networks:Fundmental Properties and Network
  Topologies}} (\bibinfo{publisher}{Springer},
  \bibinfo{address}{Berlin, Germany}, \bibinfo{year}{2006}).

\bibitem[{\citenamefont{Senthilkumar and
  Lakshmanan}(2005{\natexlab{a}})}]{dvskml2005}
\bibinfo{author}{\bibfnamefont{D.~V.} \bibnamefont{Senthilkumar}}
  \bibnamefont{and}
  \bibinfo{author}{\bibfnamefont{M.}~\bibnamefont{Lakshmanan}},
  \bibinfo{journal}{Phys. Rev. E} \textbf{\bibinfo{volume}{71}},
  \bibinfo{pages}{016211} (\bibinfo{year}{2005}{\natexlab{a}});
\bibinfo{journal}{Phys. Rev. E} \textbf{\bibinfo{volume}{76}},
  \bibinfo{pages}{066210} (\bibinfo{year}{2007}{\natexlab{a}}).

\bibitem[{\citenamefont{Senthilkumar and
  Lakshmanan}(2005{\natexlab{a}})}]{dvskjk2009}
\bibinfo{author}{\bibfnamefont{D.~V.} \bibnamefont{Senthilkumar}},
\bibinfo{author}{\bibfnamefont{J.} \bibnamefont{Kurths}}
  \bibnamefont{and}
  \bibinfo{author}{\bibfnamefont{M.}~\bibnamefont{Lakshmanan}},
\bibinfo{journal}{Chaos} \textbf{\bibinfo{volume}{19}},
  \bibinfo{pages}{023107} (\bibinfo{year}{2009}).

\bibitem[{\citenamefont{Senthilkumar and
  Lakshmanan}(2005{\natexlab{b}})}]{dvskmlijbc}
\bibinfo{author}{\bibfnamefont{D.~V.} \bibnamefont{Senthilkumar}}
  \bibnamefont{and}
  \bibinfo{author}{\bibfnamefont{M.}~\bibnamefont{Lakshmanan}},
  \bibinfo{journal}{Int. J. Bifurcation and Chaos}
  \textbf{\bibinfo{volume}{15}}, \bibinfo{pages}{2895}
  (\bibinfo{year}{2005}{\natexlab{b}});
\bibinfo{author}{\bibfnamefont{P.}~\bibnamefont{Thangavel}},
  \bibinfo{author}{\bibfnamefont{K.}~\bibnamefont{Murali}}, \bibnamefont{and}
  \bibinfo{author}{\bibfnamefont{M.}~\bibnamefont{Lakshmanan}},
  \bibinfo{journal}{Int. J. Bifurcation and Chaos}
  \textbf{\bibinfo{volume}{8}}, \bibinfo{pages}{2481} (\bibinfo{year}{1998}).

\bibitem[{\citenamefont{Senthilkumar et~al.}(2006)\citenamefont{Senthilkumar,
  Lakshmanan, and Kurths}}]{dvskml2006}
\bibinfo{author}{\bibfnamefont{D.~V.} \bibnamefont{Senthilkumar}},
  \bibinfo{author}{\bibfnamefont{M.}~\bibnamefont{Lakshmanan}},
  \bibnamefont{and} \bibinfo{author}{\bibfnamefont{J.}~\bibnamefont{Kurths}},
  \bibinfo{journal}{Phys. Rev. E} \textbf{\bibinfo{volume}{74}},
  \bibinfo{pages}{035205(R)} (\bibinfo{year}{2006});
  \bibinfo{journal}{Chaos} \textbf{\bibinfo{volume}{18}},
  \bibinfo{pages}{023118} (\bibinfo{year}{2008}).

\bibitem[{\citenamefont{Boccaletti et~al.}(2001)\citenamefont{Chen,
  and Kurths}}]{mcjk2007}
\bibinfo{author}{\bibfnamefont{M.}~\bibnamefont{Chen}},
  \bibnamefont{and} \bibinfo{author}{\bibfnamefont{J.}~\bibnamefont{Kurths}},
  \bibinfo{journal}{Phys. Rev. E} \textbf{\bibinfo{volume}{76}},
  \bibinfo{pages}{036212} (\bibinfo{year}{2007}).

\bibitem[{\citenamefont{Boccaletti et~al.}(2000)\citenamefont{Zhou,
  Li, and Wu}}]{mz2000}
\bibinfo{author}{\bibfnamefont{M.}~\bibnamefont{Zochowski}},
  \bibinfo{journal}{Physica D} \textbf{\bibinfo{volume}{145}},
  \bibinfo{pages}{181} (\bibinfo{year}{2000});
\bibinfo{author}{\bibfnamefont{C. D.}~\bibnamefont{Li}},
\bibinfo{author}{\bibfnamefont{X. F.}~\bibnamefont{Liao}},
  \bibnamefont{and} \bibinfo{author}{\bibfnamefont{T. W.}~\bibnamefont{Huang}},
  \bibinfo{journal}{Chaos} \textbf{\bibinfo{volume}{17}},
  \bibinfo{pages}{013103} (\bibinfo{year}{2007});
\bibinfo{author}{\bibfnamefont{T. W.}~\bibnamefont{Huang}},
\bibinfo{author}{\bibfnamefont{C. D.}~\bibnamefont{Li}},
  \bibnamefont{and} \bibinfo{author}{\bibfnamefont{X. Z.}~\bibnamefont{Liu}},
  \bibinfo{journal}{Chaos} \textbf{\bibinfo{volume}{18}},
  \bibinfo{pages}{033122} (\bibinfo{year}{2008}).

\end{thebibliography}

\end{document}